\documentclass[12pt,letterpaper]{article}

\usepackage{amsmath,amssymb,calc}
\usepackage{graphicx}

\newcommand\be{\begin{equation}}
\newcommand\ee{\end{equation}}
\newcommand{\bea}{\begin{eqnarray}}
\newcommand{\eea}{\end{eqnarray}}

\newcommand{\nn}{\nonumber}
\newcommand{\pd}{\partial}

\def\id{\protect{{1 \kern-.28em {\rm l}}}}

\def\id{\protect{{1 \kern-.28em {\rm l}}}}

\begin{document}

\begin{titlepage}
\begin{center}
\hfill QMUL-PH-06-08 \\
\hfill {\tt hep-th/0606271}\\
\vskip 15mm

{\Large {\bf Five-brane Instantons vs Flux-induced Gauging of Isometries \\[3mm] }}

\vskip 10mm

{\bf Lilia Anguelova and Konstantinos Zoubos}

\vskip 4mm
{\em Department of Physics, Queen Mary, University of London}\\ 
{\em Mile End Road, London E1 4NS, United Kingdom} \\
{\tt l.anguelova@qmul.ac.uk, k.zoubos@qmul.ac.uk}\\

\vskip 6mm

\end{center}

\vskip .1in

\begin{center} {\bf ABSTRACT }\end{center}

\begin{quotation}\noindent

In five-dimensional heterotic M-theory there is necessarily nonzero background flux, which 
leads to gauging of an isometry of the universal hypermultiplet moduli space. This isometry, however, is poised to be broken by M5-brane instanton effects. We show that, 
similarly to string theory, the background flux allows only brane instantons that preserve 
the above isometry. The zero-mode counting for the M5 instantons is related to the number 
of solutions of the Dirac equation on their worldvolume. We investigate that equation in 
the presence of generic background flux and also, in a particular case, with nonzero 
worldvolume flux.

\end{quotation}
\vfill

\end{titlepage}

\eject

\tableofcontents

\section{Introduction}

Since the realization that nonzero background fluxes play an essential role for 
the resolution of the moduli stabilization problem in string and M- theory \cite{DRS,GKP}, there has been an
enormous amount of work on various flux compactifications. For a comprehensive review see 
\cite{Grana}. The low-energy description of these compactifications is given by 
supergravity coupled to a certain number of matter and vector multiplets. The scalars 
belonging to these multiplets parametrize the moduli space of the four-dimensional effective 
theory. Depending on the preserved amount of supersymmetry, the presence of flux leads either to generation of a 
superpotential for (some of) the moduli fields or to gauging of some of the moduli 
space isometries.

These isometries survive perturbative corrections, but not necessarily non-perturbative 
ones. A source of the latter kind of contributions to the moduli space metric in $N=2$ 
compactifications is provided by Euclidean branes.\footnote{These nonperturbative corrections are 
essential for the stabilization of the K\"{a}hler moduli in type IIB compactifications 
\cite{KKLT}.} Because of charge quantization, these 
brane instantons lead to the breaking (to a discrete subgroup) of certain continuous isometries. The latter are 
the shift symmetries implied by the gauge invariance of the 10d or 11d 
$p$-forms that couple to those branes. Since only continuous isometries can be gauged, 
there seems to be a potential clash between turning on background fluxes and taking into 
account brane-instanton effects. The resolution of this problem for string theory was 
addressed in \cite{KT}. It was shown there that, in the presence of D0- and D2-brane 
instantons, the background fluxes protect exactly the isometries that need to be gauged 
thus preserving the consistency of the supergravity description. 

We will generalize their argument for the case of five-dimensional heterotic M-theory. A 
notable feature of the latter is that, unlike string theory, it does not admit vanishing 
background flux and so it always has a gauged  isometry. This theory arises from 
considering Ho\v{r}ava-Witten on a CY(3) and is of interest for the following reason. The 
Ho\v{r}ava-Witten set-up is given by M-theory on an 11d manifold with boundaries 
or, equivalently, compactification of M-theory on an interval. To get to four dimensions, one further compactifies on a 
CY(3). This provides the strong coupling description of the $E_8\times E_8$ heterotic string compactification 
on the same Calabi-Yau \cite{HW} and was argued to improve significantly on the 
phenomenological properties of the weakly coupled limit \cite{EW}. However, comparing to 
phenomenology leads to the conclusion that at some range of high energies the size of the 
interval is significantly larger than the size of the Calabi-Yau \cite{BD}. There is then 
an energy range or equivalently a time interval during the early universe evolution, when 
the universe is effectively five-dimensional.\footnote{At even earlier times (or higher 
energies) it is eleven-dimensional, whereas at later times (or lower energies) it becomes 
four-dimensional.} The implications of this picture for cosmology have been studied 
extensively in recent years \cite{cosm}.

The effective action of 5d heterotic M-theory was obtained in 
\cite{LOSW_DW,LOSW}. It is given by 
five-dimensional $N=2$ gauged supergravity coupled to a certain number of vector- and 
hypermultiplets. Among them is a hypermultiplet that is the same for every Calabi-Yau and 
is hence called universal. The isometry that is gauged due to the presence of background 
flux is a symmetry of the universal hypermultiplet moduli space. On the other hand, one 
can show that this is precisely the symmetry that gets broken by the presence of M5-brane 
instantons wrapping the entire CY(3). (These instantons play a crucial role for moduli stabilization in 4d heterotic M-theory \cite{CKM5}.) The reasoning is in exact parallel with the 
considerations of \cite{BB}, which related 2- and 5-brane instantons to the breaking of 
particular isometries of the universal hypermultiplet target space in type IIA string 
theory on a CY(3). We will see that the apparent contradiction can be resolved along the 
lines of \cite{KT}. This leads to certain topological restrictions on the Calabi-Yau 
three-fold in order for M5 instantons to be present, but these restrictions can be eased by 
considering compactifications with non-standard embedding.

However, one could ask whether the M5 instantons contribute to the metric at all since a 
background flux can lift (some of) the fermionic zero modes living on the M5 worldvolume 
\cite{KKT, NS, BKKST}.\footnote{This effect was first anticipated in \cite{GKTT}.} 
The latter works studied the influence of nonvanishing flux on the zero mode counting 
in the context of non-perturbatively generated superpotentials in $N=1$ 
compactifications. Although it has been known for a while \cite{BBSHM} 
that brane instantons can contribute to 
the superpotential, still many conceptual issues about the computation 
of their effects remain open. In the particular case of an M5-brane instanton 
wrapping a codimension-two cycle $D$ of the internal space in an M-theory compactification 
on a CY(4), it was shown in \cite{WNonpSup}, in the absence of flux, that the 
instanton gives a nonvanishing contribution only when the arithmetic genus $\chi (D, 
{\cal O}_D)$ of the cycle is equal to 1. This condition is required in order to cancel the 
$U(1)$ anomaly related to rotations in the two internal dimensions that are normal to the 
M5 world-volume. In this compactification one obtains an exact result for the superpotential 
and this result can be translated into an exact superpotential for some particular cases 
in type IIB or heterotic compactifications via dualities.\footnote{The relation with IIB 
comes via F-theory when the CY(4) is an elliptic fibration over a complex three-fold $X$. 
Then, if $X$ is itself a ${\bf P}^1$ fibration over a two-fold $Y$, one can map to the 
heterotic string on a $T^2$ fibration over $Y$ \cite{Vafa}.} The recent developments regarding 
zero-mode counting on brane-instanton world-volumes, that are based on the Dirac 
equation derived in \cite{KS}, all address the issue of how the requirement $\chi (D, 
{\cal O}_D) = 1$ changes in the presence of background flux. However, even when the 
brane-instanton wraps the entire internal space, as is the case with the M5-instanton in 
5d heterotic M-theory, and so there is no $U(1)$ anomaly to be considered, still the 
supersymmetries that are broken by the brane-instanton generate fermionic zero modes in 
its world-volume theory. Hence studying the Dirac equation on the M5 world-volume can tell 
us when the brane-instanton can contribute to the moduli space metric. 

As is well-known \cite{EW}, the supersymmetric backgrounds in heterotic M-theory can have 
the following nonvanishing components of the 11d supergravity four-form $G$: $(2,2,0)$, 
$(2,1,1)$ and $(1,2,1)$, where the first two digits are the number of legs along the 
holomorphic and antiholomorphic indices of the Calabi-Yau three-fold respectively and the 
last one is along the interval direction. We will see that fluxes of type $(2,2,0)$ and 
$(2,1,1)$ do not affect the zero-mode counting on the M5 world-volume thus leading to four 
zero modes as in the fluxless case, whereas flux of type $(1,2,1)$ lifts all fermionic zero 
modes. We interpret this to mean that M5 instantons are incompatible with $(1,2,1)$ flux 
backgrounds. For anti-M5 instantons the situation is reversed, i.e. it is the $(2,1,1)$ 
type of flux that lifts all their zero modes.

Finally, we address the role of the self-dual three-form, living on the M5 brane, for the 
zero-mode counting of the world-volume fermions. This field has always been neglected in 
the literature because its presence complicates the Dirac equation quite a lot. However, 
it is a crucial ingredient in the generalization of the arguments of \cite{KT} to our case.
So it is natural to ask how it would affect the above considerations. We do not undertake 
an investigation of the most general situation either, but for a particular case we are able 
to solve the Dirac equation for the most generic world-volume flux allowed by the M5 field 
equations. It turns out, that in this case the world-volume flux does not affect the zero 
mode counting.

The present paper is organized as follows. In Section \ref{Gi} we review necessary 
background material about 5d heterotic M-theory and its $(2,2,0)$-flux induced gauged 
isometry. In Section \ref{5inst} we summarize the results of \cite{BB} on the breaking 
of isometries of the universal hypermultiplet by 2- and 5-brane instantons and explain how 
this translates to 5d heterotic M-theory. In Section \ref{reconc} we tackle the 
reconciliation of M5-brane instantons with the gauged isometry of heterotic M-theory. In 
\ref{Pr} we show that the M5 instantons can indeed contribute to the moduli space metric on 
the basis of the zero mode counting on their world-volume. In \ref{Res} we argue that, 
similarly to the case considered in \cite{KT}, the Gauss's law on the M5 world-volume 
forbids exactly the instantons that would have broken the gauged isometry. The existence 
of M5 instantons, which do not break this isometry, is related to a topological restriction 
on the internal CY(3). In \ref{NonSE} we show that this restriction can be eased in 
compactifications with non-standard embeddings. In Section \ref{Otherfl} we consider other 
types of background flux and show that the zero mode counting of the M5 world-volume Dirac 
equation is not affected by $(2,1,1)$ flux, whereas all zero modes are lifted by $(1,2,1)$ 
flux. In the Appendix we show that the roles of these two types of flux are reversed for anti-M5 brane instantons. Finally, in Section \ref{WVfl} we consider in a particular case the Dirac equation 
with nonvanishing world volume flux and find that the latter does not change the zero mode 
counting.

\section{Gauged isometry} \label{Gi}

The effective five-dimensional theory arising from compactification of Ho\v{r}ava-Witten on a
CY(3) was considered in \cite{LOSW}. It was shown there, that this is gauged supergravity 
coupled to $h^{1,1}-1$ vector multiplets and $h^{2,1}+1$ hypermultiplets. The $+1$ is 
the universal hypermultiplet that appears for any CY(3). Its bosonic field content is the 
following: the CY volume $V$, a real scalar $\sigma$ that is dual to the external 
components of the 11d supergravity 3-form $C$ and a complex scalar $\xi$ which comes from 
$C = \xi \Omega +...$, where $\Omega$ is the holomorphic 3-form of the CY space. The 
presence of boundaries in eleven dimensions leads to a modification of the Bianchi 
identity for the field strength $G$ of $C$: 
\be \label{BI}
dG = -\frac{1}{2\sqrt{2}\pi} \,\left(\frac{\kappa}{4\pi}\right)^{2/3} \,\sum_{a=1}^2 
\delta(x^{11} - x^{(a)}) \,\left( {\rm tr} F^{(a)}\wedge F^{(a)} - \frac{1}{2} {\rm tr} 
R\wedge R \right) \, ,
\ee
where $x^{(1)} = 0$ and $x^{(2)} = \pi \rho$ are the positions of the two boundaries. 
As a result, only solutions with nonzero background flux are allowed. That  
is precisely the reason for the gauging of the effective 5d supergravity. This 
gauging will be important in the following. So, in order to explain how it occurs, let us first 
introduce the relevant notation and conventions of \cite{LOSW}.

Let us start by taking the standard embedding of the spin connection in the first 
gauge group:\footnote{The other $E_8$ gauge bundle is taken to be trivial.}
\be \label{StEmb}
{\rm tr}F^{(1)}\wedge F^{(1)} = {\rm tr} R\wedge R \, .
\ee
Then (\ref{BI}) becomes:
\be \label{BI1}
(dG)_{11ABCD} = -\frac{1}{4\sqrt{2}\pi}\left(\frac{\kappa}{4\pi}\right)^{2/3} \left[\delta
(x^{11}) - \delta (x^{11} - \pi \rho)\right] ({\rm tr} R\wedge R)_{ABCD} \, ,
\ee
where the indices $A,B,C,D$ run over the six CY directions. Following \cite{LOSW}, we 
introduce a basis $\nu^i$, $i=1,..,h^{2,2}=h^{1,1}$, of $(2,2)$-forms on the CY such that:
\be
\frac{1}{v^{2/3}} \int_{C_i} \nu^j = \delta_{\,i}^j \, , \qquad \frac{1}{v} \int_X 
\nu^{\,i} \wedge \omega_j = \delta_j^{\,i} \, ,
\ee
where $C_i$ is a basis of 4-cycles, $\omega_j$ $-$ a basis of $(1,1)$ forms and $v$ is a 
6d reference volume. Now, one can expand the non-exact part of ${\rm tr}R \wedge R$ as:\footnote{In the language of \cite{LOSW} this is denoted as ${\rm tr} R \wedge R|_0$ and 
referred to as `zero mode part'. Its existence is exactly what leads to their 
$G_{ABCD}|_0 \neq 0$, which in modern terminology is really the background flux.}
\be \label{RRne}
{\rm tr}R \wedge R|_{ne} = -8 \sqrt{2} \pi \left( \frac{4\pi}{\kappa}\right)^{2/3} \alpha_i 
\nu^i \, .
\ee
The numerical coefficient above is chosen for convenience and
\be \label{al}
\alpha_i = \frac{\pi}{\sqrt{2}} \left( \frac{\kappa}{4\pi} \right)^{2/3} \frac{1}{v^{2/3}} 
\,\beta_i \, , \qquad \beta_i = - \frac{1}{8\pi^2} \int_{C_i} {\rm tr} R\wedge R
\ee
with $\beta_i$ being integers related to the first Pontrjagin class of the CY. Using (\ref{RRne}), the Bianchi
identity (\ref{BI1}) and the field equation, $D_I G^{IJKL}=0$ with $I=1,...,11$ \,, \,can 
be solved by the following background flux:
\bea \label{BcGr}
G_{ABCD} &=& \alpha_i \nu^{\,i}_{ABCD} \epsilon(x^{11}) \, , \nn \\
G_{ABC11} &=& 0 \, ,
\eea
where $\epsilon(x^{11})$ is the step function defined to be $+1$ for $x^{11}>0$ and $-1$ for $x^{11}<0$.

Now we are ready to state the result of \cite{LOSW} about the flux-induced gauging of an 
isometry of the universal hypermultiplet moduli space. Let us denote the coordinates on 
the latter by $q^u\equiv (V, \sigma, \xi, \bar{\xi})^u$. Then the kinetic term of the 
universal hypermultiplet is \cite{LOSW}:
\be \label{GIs}
h_{uv} D_{\alpha} q^u D^{\alpha} q^v \, , \qquad D_{\alpha} q = (\pd_{\alpha} V, 
\,\pd_{\alpha} \sigma - 2 \epsilon(x^{11}) \alpha_i {\cal A}^i_{\alpha}, \,\pd_{\alpha} 
\xi, \,\pd_{\alpha} \bar{\xi}) \, ,
\ee
where $h_{uv}$ is the metric on the quaternionic space $SU(2,1)/U(2)$ and 
${\cal A}^i_{\alpha}$ are $h^{1,1}$ gauge fields arising via $C_{\alpha AB} = \frac{1}{6} 
{\cal A}^i_{\alpha} \omega_{i AB}$ with the index $\alpha$ running along the five non-CY 
dimensions\footnote{The number of vector multiplets is $h^{1,1}-1$ because one of the 
$h^{1,1}$ vectors ${\cal A}^i$ (rather, a certain combination of them) is the graviphoton 
of the supergravity multiplet.}. Clearly, the isometry $\sigma \rightarrow \sigma + const$ 
of the metric $h_{uv}$ is now gauged because of the nonzero background flux 
$G = \alpha_i \nu^i \,\epsilon(x^{11})$. The dualization that relates $\sigma$ and 
$G_{\alpha \beta \gamma \delta}$ is accordingly modified:
\be \label{Gs}
G = \frac{1}{\sqrt{2}} V^{-2} *_5 \!\left[d\sigma -2\epsilon(x^{11}) \alpha_i {\cal A}^i 
- i(\xi d\bar{\xi} - \bar{\xi} d\xi) \right] .
\ee
Finally, comparing (\ref{GIs}) with the general expression for the extended derivative, 
$D_{\alpha} q^u = \pd_{\alpha} q^u + g {\cal A}_{\alpha}^i k_i^u$, we see that the Killing 
vectors $k_i$ are:\footnote{For convenience we set the gauge coupling constant $g=1$; or, equivalently, 
we absorb it in the definition of $\alpha_i$.}
\be \label{ki}
k_i = -2 \epsilon(x^{11}) \alpha_i \pd_{\sigma} \, .
\ee

Although defining $k_i$ as above will be of use for us, we should note that strictly speaking there is only one Killing vector: $k=\pd_{\sigma}$. So, in fact, the 
gauge field for the gauging is a linear combination of the graviphoton and the vectors 
from the vector multiplets, which is given by ${\cal A}_{\alpha}= -2 \epsilon(x^{11}) \alpha_i 
{\cal A}^i_{\alpha}$.

\section{Five-brane instantons} \label{5inst}
\setcounter{equation}{0}

In \cite{BB} it was shown that 5-brane and 2-brane instantons lead to the breaking of 
certain isometries of the universal hypermultiplet moduli space. The considerations of 
that paper were in the context of type IIA compactifications on a CY(3) to a 
four-dimensional effective theory with $N=2$ supersymmetry. In this case the bosonic 
content of the universal hypermultiplet is made up of the dilaton $\varphi$, a real scalar 
$D$ that is dual to the external components of the B-field and a complex scalar $C$ 
originating from $A_{(3)} = C \Omega$, where $A_{(3)}$ is the RR 3-form potential. The 
precise relation between $D$ and $H$, which is locally $H_{\mu \nu \rho} = (dB)_{\mu \nu 
\rho}$ with $\mu, \nu, \rho$ being four-dimensional indices, is given by
\be \label{BD}
H = e^{4 \varphi} *_4 \!\left[2 dD + i (\bar{C} dC - C d\bar{C})\right] .
\ee
The manifold parametrized by $\varphi, D, C, \bar{C}$ is the coset $SU(2,1)/U(2)$ \cite{FS}. 
In terms of the complex coordinates $S, \bar{S}, C, \bar{C}$, where
\be \label{CoordCh}
S \equiv e^{-2\varphi} + 2iD + C \bar{C} \, ,
\ee
this coset has the following symmetries:
\bea \label{sym}
S &\rightarrow& S + i \alpha + 2 (\gamma + i \beta) C + \gamma^2 + \beta^2 \nn \\
C &\rightarrow& C + \gamma - i \beta \, ,
\eea
which correspond to constant shifts of the NS axion $D$ and the RR scalars $C, \bar{C}$. 
These symmetries are invariances of the classical Lagrangian. Their existence is implied by the gauge transformations\footnote{Recall that gauge invariance implies lack of non-derivative couplings of the corresponding potentials, which in turn gives rise to exactly the shift symmetries of the scalars obtained from reduction of those potentials.} of the 3-form $H$ and 4-form $F_{(4)}=dA_{(3)}$. As shown in \cite{EWNewIss}, they survive when sigma-model perturbative corrections are taken into account. They are also expected to survive in string perturbation theory \cite{AStr}.\footnote{The 
perturbative corrections to the moduli space of the universal hypermultiplet were first 
addressed in \cite{AStr} and studied more thoroughly in \cite{AMTV, ARV}.} (By contrast, the 
remaining symmetries of the coset $SU(2,1)/U(2)$ are generically broken by perturbative 
effects.\footnote{For more details on the symmetries of the coset $SU(2,1)/U(2)$ see 
\cite{dWvP}.}) However, non-perturbative corrections due to membrane and five-brane 
instantons will break the isometries in (\ref{sym}) \cite{BB}. Let us recall the argument 
for this.

It was shown in \cite{BB} that the symmetries (\ref{sym}) give rise to the No\"{e}ther 
currents
\bea
J_{\alpha} &=& \frac{i}{\kappa^2_4} e^{2K} \left( dS - d\bar{S} + 2 (C d\bar{C} - 2\bar{C} 
dC) \right) \, ,\nn \\
J_{\beta} &=& -\frac{2i}{\kappa^2_4} e^K (dC - d\bar{C}) + 2 (C + \bar{C}) J_{\alpha} \, 
,\nn \\
J_{\gamma} &=& -\frac{2}{\kappa^2_4} e^K (dC + d\bar{C}) - 2i (C - \bar{C}) J_{\alpha} \, 
,
\eea
where $K$ is the K\"{a}hler potential $K = - \ln (S + \bar{S} - 2 C \bar{C})$. Integrating 
these currents over a three-cycle $\Sigma_3$ in the 4d external space, one obtains the 
corresponding conserved charges:
\be
Q_{\alpha, \beta, \gamma} = \int_{\Sigma_3} *_4 \,J_{\alpha, \beta, \gamma} \, .
\ee
However, these charges can be shown to be related to the presence of 5-brane ($Q_{\alpha}$) 
and 2-brane ($Q_{\beta}$, $Q_{\gamma}$) instantons. For example for the 5-brane, the case 
that will be of importance for us, one can easily see from (\ref{CoordCh}) and (\ref{BD}) 
that 
\be
Q_{\alpha} = \int_{\Sigma_3} *_4 J_{\alpha} = \int_{\Sigma_3} H \, ,
\ee
where we also used that $K = - \ln (S + \bar{S} - 2 C \bar{C}) = 2 \varphi$. Clearly then, 
$Q_{\alpha}$ is the five-brane charge and so charge quantization implies that the presence 
of 5-brane instantons breaks the symmetry generated by $J_{\alpha}$ (i.e. the symmetry 
$S \rightarrow S + i \alpha$) to a discrete subgroup.\footnote{For further study of 2- and 5-brane instanton effects on the universal hypermultiplet moduli space in the supergravity description see \cite{SVandoren}.}

Let us now compare the above type IIA compactification to a 4d $N=2$ theory with the 
compactification of Ho\v{r}ava-Witten on CY(3), that leads to a five-dimensional effective 
theory. In both cases there are eight preserved supercharges. In addition, the scalars 
$V, \sigma, \xi, \bar{\xi}$, introduced in the previous section, parametrize the same 
quaternionic manifold, $SU(2,1)/U(2)$, as do $\varphi, D, C, \bar{C}$. The coordinate 
transformation between the two sets of coordinates is:
\be
\sigma = 2D \, , \qquad V = e^{-2\varphi} \, , \qquad \xi = C \, .
\ee
Hence, the same symmetries as (\ref{sym}) are also present for the moduli space of the 
universal hypermultiplet in the 5d theory. And similarly to the type IIA case this leads, 
upon using (\ref{Gs}), to the shift symmetry $\sigma \rightarrow \sigma + \alpha$ being 
broken by the presence of five-brane charge\footnote{Since $D= Im S$, clearly the shift 
$S \rightarrow S + i \alpha$ is actually $D \rightarrow D + \alpha$.}
\be
Q_{\alpha} = \int_{\Sigma_4} *_5 J_{\alpha} = \int_{\Sigma_4} G \, ,
\ee
where $\Sigma_4$ is a four-cycle in the five-dimensional external space.\footnote{A 
similar conclusion was reached in \cite{GS} from a different point of view.}

However, as we recalled in Section \ref{Gi}, the isometry $\sigma \rightarrow \sigma + 
const$ is gauged because of the presence of background flux in the CY compactification of 
Ho\v{r}ava-Witten theory. Since only continuous isometries can be gauged, it appears therefore 
that there is a clash between this gauging and the possible five-brane instanton effects.

\section{Flux-induced gauging vs M5 instantons} \label{reconc}

In the present section we address the reconciliation of the above competing effects. First,
in \ref{Pr} we explain that five-brane instantons can exist in the theory we are 
considering and so there is indeed a potential problem. In \ref{Res} we show that the 
latter is resolved, similarly to the string theory case, by requiring that Gauss' law is 
obeyed on the brane-instanton world-volume. This implies that the background flux allows 
only instantons that would not break the gauged isometry. If such M5's are to exist, then
the CY has to satisfy some topological constraints. In \ref{NonSE} we show that these 
constraints can be made less restrictive by considering compactifications with 
non-standard embedding due to the presence of Minkowski M5-branes in the bulk.

\subsection{Fermionic zero modes and 5-brane instantons} \label{Pr}

To claim that there is a possible clash between the gauged isometry, parametrized by the 
coordinate $\sigma$, and five-brane instantons, let us first convince ourselves that the 
latter are not forbidden by supersymmetry. In \cite{LLO} it was shown that, to first order 
in the $\kappa^{2/3}$ expansion of Ho\v{r}ava-Witten theory, supersymmetry allows only 
Minkowski membranes that stretch between the two boundaries and Minkowski five-branes that 
are parallel to the boundaries. These are solutions in which the M2 and M5 branes are part 
of the background. Nevertheless, it is natural to expect that the same conclusion will 
hold for their instantonic counterparts. Indeed, it was shown in \cite{LOPR} that the only 
M2 instantons, which are compatible with supersymmetry, are given by membranes wrapping 
holomorphic curves on the boundaries and stretching along the eleventh direction. The 
argument was based on analyzing what embeddings of the membrane worldvolume into the 
eleven-dimensional spacetime allow solutions of $\Gamma \epsilon = \epsilon$, where 
$\Gamma$ is the worldvolume operator that defines the $\kappa$-symmetry transformation and 
$\epsilon$ is the supersymmetry parameter. Clearly, one can perform an analogous 
computation for the 5-brane instantons. However, it will be of future use for us to verify 
the existence of instantons, due to M5-branes wrapping the whole CY(3), by counting the 
fermionic zero-modes on the brane worldvolume.

Recall that the supersymmetries that are broken by the presence of a brane generate 
fermionic zero modes on its worldvolume.\footnote{For a nice recent discussion of this 
point see \cite{BKKST}.} In order for a brane-instanton to be able to contribute to the 
moduli space metric of the external theory, the Dirac equation for its worldvolume fermions 
has to have four zero modes. Note that, in addition to the zero modes coming from the 
broken supercharges, there can also be zero modes related to internal degrees of freedom 
(i.e., superpartners of bosonic deformations of the internal cycle). Furthermore, recently 
it was shown that background fluxes can change the zero-mode counting significantly 
\cite{KKT,NS}.\footnote{An apparent contradiction in this kind of analysis, related to the consistent inclusion of non-perturbative effects in the minimization of the 4d effective superpotential, was resolved in \cite{LRST}.} These results were based on the Dirac equation for the M5-worldvolume 
fermions in a nonvanishing background, derived in \cite{KS}. The latter work considers 
only the quadratic terms in the fermionic worldvolume action. However, this is enough for 
ruling out M5-brane instanton  contributions (in the case of less than four zero modes) 
since the higher (interaction) terms can only lift zero modes of the quadratic action but 
not introduce new ones.\footnote{Recall that, in principle, counting arguments can only be 
enough for ruling out certain contributions. However, they do not necessarily imply a 
non-vanishing correction since even when they do not rule it out, the explicit computation 
may end up giving zero.} As we reviewed in Section \ref{Gi}, in the case of interest for 
us there is nonvanishing background flux. So we are going to show that there are exactly 
four zero modes on the worldvolume of an M5 wrapping a CY 3-fold by specializing the Dirac 
equation of \cite{KS} to our set-up.

Let us start by decomposing the eleven-dimensional spinor in the appropriate way. To 
begin with, it transforms in the ${\bf 32}$ of $SO(1,10)$, or in $Spin(1,10)$ to be more precise. Compactifying on 
$CY\times M_4 \times S^1/{\mathbb Z}_2$, the group $SO(1,10)$ gets broken to 
$SU(3)\times SO(1,3)$. After analytic continuation to Euclidean space, the latter group becomes $SU(3)\times SO(4)$. Hence the spinor, $\theta$, on the worldvolume of an M5-brane wrapping the 
CY(3) transforms in the $({\bf 1},{\bf 4})\oplus ({\bf 3},{\bf 4}) \oplus 
({\bf \bar{3}},{\bf 4}) \oplus ({\bf \bar{1}}, {\bf 4})$. Defining the Clifford vacuum 
$|\Omega \rangle$ by
\be
\gamma^a |\Omega\rangle = 0 \, ,
\ee
where the index $a$ runs over the holomorphic coordinates of the CY, one can expand 
$\theta$ as
\be \label{ThExp}
\theta = \phi |\Omega \rangle + \,\, \phi_{\bar{a} \bar{b}} \gamma^{\bar{a} \bar{b}}|\Omega\rangle 
\, .
\ee 
Recall that this expansion contains only terms with even number of indices because after 
$\kappa$-symmetry gauge-fixing one is left with a chiral fermion on the worldvolume 
\cite{KS}. Also, we have suppressed the ${\bf 4}$ index on $\phi$, $\phi_{\bar{a}\bar{b}}$ 
for simplicity. 

Let us now turn to the Dirac equation \cite{KS}:
\be \label{DEq}
\gamma_A m^{AB} \nabla_B \theta + \frac{1}{24} \left[ \gamma^{\alpha \beta \delta} 
\gamma^A (2\delta_A^B - m_A{}^B) G_{B\alpha \beta \delta} + \gamma^{\alpha} \gamma^{ABC} 
(2\delta_A^D - 3m_A{}^D) G_{DBC\alpha} \right] \theta = 0 \, .
\ee
As before, indices $\alpha, \beta, \delta$ run over the five dimensions that are 
transverse to the CY (and so to the M5-brane instanton) and $A,B,C,D$ run along the six 
worldvolume directions. Also, the matrix $m$ is determined by the worldvolume 3-form flux 
$h$ via 
\be
m_A{}^B = \delta_A{}^B - 2 h_{ACD} h^{BCD} \, .
\ee
For convenience, from now on we will absorb the $1/24$ factor in the definition of the 
background flux $G$. To simplify the problem we will consider in the following, as in all
existing literature, only vanishing worldvolume flux. (We will have more to say about the 
$h\neq 0$ case in Section \ref{WVfl}.) Hence (\ref{DEq}) reduces to:
\be \label{DirEq}
\gamma^a \nabla_a \theta + \gamma^{\bar{a}} \nabla_{\bar{a}} \theta +  
\left( \gamma^{\alpha \beta \delta} \gamma^A G_{A \alpha \beta \delta} - \gamma^{\alpha} 
\gamma^{ABC} G_{ABC \alpha} \right) \theta = 0 \, .
\ee
Since the background flux in (\ref{BcGr}) has only $G_{ABCD}$ nonzero components, clearly 
the Dirac equation (\ref{DirEq}) is completely unaffected by the flux. Hence the counting 
of zero modes gives four, which is what is necessary for the M5 instanton to contribute to 
the metric. Indeed, substituting (\ref{ThExp}) in 
(\ref{DirEq}), one finds
\bea \label{De0}
\pd_{[\bar{a}} \phi_{\bar{b} \bar{c}]} \gamma^{\bar{a}\bar{b}\bar{c}} |\Omega\rangle &=& 0 
\nn \\
(\pd_{\bar{a}} \phi + 4 g^{\bar{b} c} \pd_c \phi_{\bar{b}\bar{a}}) \gamma^{\bar{a}} 
|\Omega\rangle &=& 0 \, ,
\eea
where $g_{a \bar{b}}$ is the K\"{a}hler metric on the CY. Hence, as in \cite{KKT}, the 
forms $\phi$ and $\phi_{\bar{a}\bar{b}}$ are harmonic.\footnote{Recall that this can be 
derived in the following way. Acting with $\nabla^{\bar{a}}$ on the second equation in 
(\ref{De0}), we obtain that $\Delta \phi = 0$. On the other hand, acting with 
$\nabla_{\bar{d}}$ on the second line of (\ref{De0}) and anti-symmetrizing w.r.t. the pair 
($\bar{a}$, $\bar{d}$) gives, after adding the result of the action of 
$\nabla^{\bar{a}}$ on the first line of (\ref{De0}), that $\Delta \phi_{\bar{b} 
\bar{c}} = 0$. These manipulations use the fact that on a K\"{a}hler manifold the only 
nonvanishing components of the Christoffel symbols are $\Gamma_{ab}^c$ and $\Gamma_{\bar{a}
\bar{b}}^{\bar{c}}$ and also $R_{a\bar{b}c\bar{d}}=R_{a\bar{d}c\bar{b}}=R_{c\bar{b}a\bar{d}}$.} However, as $h^{0,2} = 0$ for a CY(3), it follows that $\phi_{\bar{a}
\bar{b}} = 0$. So we are left with the single component $\phi$, which due to the suppressed 
index in the ${\bf 4}$ of $SO(4)$ means that there are exactly four zero modes. Therefore, 
M5-brane instantons are in principle allowed in the theory under consideration, despite 
the existence of the flux-induced gauging of the shift symmetry along $\sigma$ (which is 
the isometry they are supposed to break).

\subsection{Reconciling M5 instantons with flux-induced isometry gauging} \label{Res}

It turns out that the resolution of the above puzzle is along the lines of \cite{KT}, 
which considered D2-brane instantons and flux-induced gauging of isometries in type II 
strings. The idea is the following. The correction to the moduli space metric, due to 
brane-instanton effects, is of the form $T e^{S_{inst}}$, where $S_{inst}$ is the brane 
action and the prefactor $T$ is made up of one-loop determinants. Generically $T$ can 
depend on some of the moduli but not on the $p$-form ones, whose shift symmetries are 
broken by the brane-instanton (in our case, the coordinate $\sigma$), because the 
dependence on the latter is fixed by the charge of the instanton.\footnote{In the case of
brane-instanton generated superpotentials (i.e., in $N=1$ compactifications), $T$ is a 
function of the complex structure moduli but, due to holomorphy, not of the K\"{a}hler ones 
(see \cite{WNonpSup}). However, for non-perturbatively generated corrections to the moduli space 
metric, clearly there is no holomorphy and so one cannot exclude K\"{a}hler moduli 
dependence of the instanton prefactor.} In other words, the above $p$-form moduli enter 
the brane-instanton induced correction only via $S_{inst}$. Hence, it is enough to show 
that the change of $S_{inst}$, generated by the Killing vector of the isometry to be 
gauged, vanishes for brane-instantons that are compatible with the background flux (i.e., 
satisfy the appropriate Gauss' law).

Let us start by recalling the covariant Minkowski M5-brane worldvolume action 
\cite{BLNPST}:
\bea \label{Ac}
S_{M5} = &-& \int d^6 x \left( \sqrt{-det (g_{mn} + i \tilde{H}_{mn})} - 
\frac{\sqrt{-g}}{4\pd_q a \pd^q a} \pd_l a \,H^{*lmn} H_{mnp} \,\pd^p a \right) \nn \\ 
&-& \int \left( C^{(6)} + \frac{1}{2} F \wedge C^{(3)} \right) \, ,
\eea
where $m,n$ are worldvolume indices, $a(x)$ is an auxiliary field and
\be
H_{lmn} = F_{lmn} - C^{(3)}_{lmn} \, , \qquad H^{*lmn} = \frac{1}{3!\sqrt{-g}} 
\epsilon^{lmnpqr}H_{pqr} \, , \qquad \tilde{H}_{mn} = 
\frac{H^*_{mnl}\pd^l a}{\sqrt{(\pd a)^2}} \, .
\ee
Finally, $F=dA$ is the field-strength of the 5-brane worldvolume two-form field $A$. 
Recall that $F$ (or, equivalently, $H$) satisfies a non-linear self-duality condition 
and there is a non-linear 
field redefinition that relates it to a worldvolume 3-form $h$, which obeys an ordinary 
linear self-duality constraint but is not related to a potential.\footnote{More precisely, 
the relation between the two fields is $H_{\underline{lmn}} = 
(\delta_{\underline{l}}^{\underline{r}} - 2 h_{\underline{lpq}} h^{\underline{rpq}}) 
(\delta_{\underline{m}}^{\underline{s}} -2 h_{\underline{mp'q'}} h^{\underline{sp'q'}}) 
h_{\underline{nrs}}$ \cite{HS} in terms of flat indices, or equivalently $H_{lmn} = 
(m^{-1})_l{}^p h_{mnp}$ \cite{HSW}.} Let us also note that the auxiliary field $a(x)$ can 
be gauged away \cite{BLNPST} and in the gauge, in which $a(x)$ is equal to one of the 
worldvolume coordinates, the second term on the first line of (\ref{Ac}) is of the form
\be
\int (F-C^{(3)})\wedge (F-C^{(3)}) \, .
\ee
Euclidean continuation of the above action is achieved, as usual, by taking the 
worldvolume time $x^0$ to $\pm i x^0$.

Now, in order to follow the logic of \cite{KT} we want to see what is the explicit 
dependence of the five-brane action on the scalar $\sigma$ so that we can compute the 
change of $S_{M5}$ under the transformation generated by the vector field $k_i$ (see eq. 
(\ref{ki})). Since $*_5 dC^{(3)} = d\sigma$ with $C^{(3)}$ having only external (which in 
particular means non-worldvolume) indices, no terms with $C^{(3)}$ in $S_{M5}$ contribute 
$\sigma$-dependence.\footnote{We use $*_5dC^{(3)} = d\sigma$ instead of the full relation 
(\ref{Gs}), because we concentrate only on the $\sigma$ (as opposed to $\xi$) dependence 
and keep only terms linear in $\kappa^{2/3}$. (As the Killing vector is proportional to 
$\alpha_i \sim \kappa^{2/3}$, in $k_i (S_{M5})$ the terms of ${\cal O}(\kappa^{2/3})$ come 
from the part of $S_{M5}$ that is zeroth order in $\kappa^{2/3}$.)} On the other hand, the 
11d duality $*_{11}dC^{(3)} = dC^{(6)}$ descends to $C^{(6)}=\sigma w$, where $w$ is the 
CY volume form. Hence
\be
\delta S_{M5} = k_i (S_{M5}) = k_i (\int C^{(6)}) = -2 \epsilon(x^{11}) \alpha_i \, . 
\ee
So, as long as $\alpha_i \neq 0$, the action is not invariant. However, on the M5 
worldvolume, X, $dH = -\frac{1}{4} G$ \cite{HS}, where $G$ is the (pullback of the) 
background flux. Therefore, on $X$ the flux $G$ has to be cohomologically trivial. From 
(\ref{BcGr}) this implies that on $X$
\be 
\alpha_i = 0 \qquad \forall i \,\, , 
\ee
which restores the invariance of $S_{M5}$. So the background flux does not allow 
five-brane instantons, unless the CY(3) is such that
\be \label{Pontr}
\int_{C_i} {\rm tr} R\wedge R = 0
\ee
for every 4-cycle $C_i$, in which case the isometry $\sigma \rightarrow \sigma + const$ is
not gauged anyway. Clearly, the conditions (\ref{Pontr}) are satisfied for Calabi-Yau's 
with vanishing first Pontrjagin class.

\subsection{Non-standard embedding} \label{NonSE}

In (\ref{StEmb}) we assumed the standard relation between the $E_8$ gauge group of the 
visible boundary and the spin connection of the CY space. However, non-standard embeddings 
allow other (than $E_6$) unbroken gauge groups on the visible boundary and so have 
attracted a lot of phenomenological interest on their own. (They were introduced in the 
context of the weakly coupled heterotic string back in \cite{nonst_weak}.) Even richer 
possibilities for the breaking of $E_8\times E_8$ arise when one considers M5-branes 
parallel to the boundaries and situated at various positions along the interval. These 
five-branes are extending along the four external directions and wrapping a holomorphic 
curve in the CY(3). Including them is incompatible with the standard embedding. For an 
undoubtedly incomplete list of the vast literature on phenomenology of these 
compactifications, see \cite{nonstemb}.

The low-energy effective theory of the strongly coupled heterotic $E_8\times E_8$ string 
on a CY(3) with non-standard embedding (with or without five-branes) was derived in 
\cite{LOW}. Depending on the energy regime of interest it is useful to compactify either 
to four or to five dimensions. In the latter case, one again obtains five-dimensional 
gauged supergravity in the bulk. The effective theory has the same form as the one for the 
standard embedding \cite{LOSW}, but the gauging parameters $\alpha_i$ are now different 
from (\ref{al}). For non-standard embedding without five-branes:\footnote{The precise 
numerical coefficients will not be important for us, so we will omit them for clarity.}
\be
\alpha_i \sim \int_{C_i} \left( {\rm tr} F^{(1)}\wedge F^{(1)} - \frac{1}{2} {\rm tr}R
\wedge R \right) \, ,
\ee
whereas in the presence of $n$ M5-branes, positioned at $x_1,..., x_n$ along the eleventh 
dimension, the parameters $\alpha_i$ change in each interval $x_k \le x^{11} \le x_{k+1}$. 
More precisely, one finds \cite{LOW}:
\be
\alpha^{(k)}_i \sim \sum_{m=0}^k \beta_i^{(k)} \epsilon(x^{11}) \qquad {\rm for} \qquad 
x^{11}\in (x_k, x_{k+1}) \, , 
\ee
where $x_0$ and $x_{n+1}$ denote the positions of the visible and hidden boundaries 
respectively and the integers $\beta_i^{(k)} = \int_{C_i} J^{(k)}$ are topological 
invariants giving the intersection number of the $k$-th five-brane with the four-cycle 
$C_i$ for $k=1,...,n$ and $i = 1, ..., h^{2,2}$.

As in Section \ref{Gi}, the isometry of the universal hypermultiplet moduli space that is 
gauged is generated by $k = \pd_{\sigma}$. So, following the arguments of Section 
\ref{Res}, we again conclude that five-brane instantons are allowed only when the relevant 
gauging parameters $\alpha_i^{(k)}$ vanish. However, since the Minkowski M5-branes are 
themselves magnetic sources of flux, one can achieve the vanishing of $\alpha_i^{(k)}$ 
with an appropriate choice of M5-branes in the bulk without the need to impose 
(\ref{Pontr}) on the CY(3). Hence, the topological conditions that the Calabi-Yau should 
satisfy, so that there can be 5-brane instantons, are least restrictive for 
non-standard-embedding vacua with bulk M5-branes.

\section{More general background flux} \label{Otherfl}
\setcounter{equation}{0}

So far we have considered only background flux of type $(2,2,0)$, i.e. $G_{a\bar{b} c 
\bar{d}}$. As we saw, it does not affect the Dirac equation of the M5 world-volume 
fermions. However, in heterotic M-theory one could also have supersymmetric backgrounds 
with nonvanishing flux components of type $(2,1,1)$ and $(1,2,1)$, i.e. $G_{ab\bar{c} 11}$ 
and $G_{\bar{a}\bar{b} c 11}$; see, for example, \cite{CK}. Such components appear also in 
a background including the gauge five-brane considered in \cite{LLO}.\footnote{This is the 
lift to strong coupling of the heterotic string solution of \cite{AS}.} Let us now see how 
they modify the zero-mode counting for M5-brane instantons.

In a vacuum with nonzero $G_{ab\bar{c} 11}$ and $G_{\bar{a}\bar{b} c 11}$, the Dirac 
equation on the worldvolume of an M5 instanton\footnote{Not to be confused with the 
Minkowski gauge five-brane that may be a part of the background.} (still neglecting the 
worldvolume flux $h$) acquires the form:
\bea \label{121}
\left( \pd_{[\bar{a}} \phi_{\bar{b} \bar{c}]} + 4 G_{[\bar{a}\bar{b}}{}^{\bar{d}}{}_{|11|} 
\phi_{\bar{c}] \bar{d}} + 2 G^{\bar{d}}{}_{\bar{d}\,[\bar{a}\,|11|} \phi_{\bar{b}\bar{c}]} \right) 
\gamma^{\bar{a}\bar{b}\bar{c}}|\Omega\rangle &=& 0 \nn \\
\left( \pd_{\bar{a}} \phi + 4 g^{\bar{b} c} \pd_c \phi_{\bar{b} \bar{a}} - 
8 G_{\bar{a} 11 bc} \phi^{bc} + 8 G_{c 11}{}^c{}^{\bar{b}} \phi_{\bar{b} \bar{a}} + 
2 G_{\bar{a}\bar{c}}{}^{\bar{c}}{}_{11}\phi \right) 
\gamma^{\bar{a}} | \Omega\rangle &=& 0 \, ,
\eea
where we have used, as before, the decomposition (\ref{ThExp}). At first sight, equations (\ref{121}) look quite complicated. However, their analysis can be facilitated by the following observations. Since they are linear in the flux, one can study the contributions of the $(2,1,1)$ and $(1,2,1)$ components separately. Furthermore, on physical grounds turning on background flux can only reduce the number of zero modes compared to the fluxless case.\footnote{The reason is that for nonzero flux there are additional supersymmetry constraints (for example, primitivity conditions for flux components). Satisfying them leads to smaller number of geometric moduli and hence also to smaller number of their superpartners, which are the fermionic moduli.} However, the presence of flux can deform (some of) the surviving zero modes. Let us see what do the above considerations imply in our case. For vanishing flux the four zero modes (recall that for convenience we have suppressed the {\bf 4} index of $\phi$, $\phi_{\bar{a}\bar{b}}$) were given by $\phi$ -- harmonic and $\phi_{\bar{a}\bar{b}} = 0$. Hence, if it is possible to have $\phi_{\bar{a}\bar{b}}\neq 0$ for $G\neq 0$, then the $\phi_{\bar{a}\bar{b}}$ solution must be completely determined by the solution for $\phi$, together with the flux. Otherwise the number of zero modes will increase by $3\times 4$, which is the number of independent components of $\phi_{\bar{a}\bar{b}}$.

Now we are ready to start analyzing the system (\ref{121}) in the presence of each of the two allowed types of flux. Let us begin with $(2,1,1)$ fluxes. In this case the equations become:
\bea \label{211}
\pd_{[\bar{a}} \phi_{\bar{b} \bar{c}]} &=& 0 \nn \\
\pd_{\bar{a}} \phi + 4 g^{\bar{b} c} \pd_c \phi_{\bar{b} \bar{a}} - 
8 G_{\bar{a} 11 bc} \phi^{bc} + 8 G_{c 11}{}^c{}^{\bar{b}} \phi_{\bar{b} \bar{a}} &=& 0 \, .
\eea
Acting on the second equation with $\nabla_{\bar{d}}$, together with antisymmetrizing w.r.t. $\bar{d}$ and $\bar{a}$, and adding to the result the action of $\nabla^{\bar{a}}$ on the first equation, we find:\footnote{Recall that $\Delta \phi_{\bar{a}\bar{b}} = 2 \Delta_{\bar{\pd}} \, \phi_{\bar{a}\bar{b}}= -2 \nabla^{\bar{c}} \nabla_{\bar{c}} \, \phi_{\bar{a}\bar{b}} - 4 R_{\bar{a}}{}^{\bar{c}}{}_{\bar{b}}{}^{\bar{d}} \, \phi_{\bar{c}\bar{d}} = -2 \nabla^{\bar{c}} \nabla_{\bar{c}} \, \phi_{\bar{a}\bar{b}}$, where the last equality is due to $R_{a\bar{b}c\bar{d}} = R_{a\bar{d}c\bar{b}}$ on a K\"{a}hler manifold.}
\be \label{DelPhi}
\Delta \phi_{\bar{d}\bar{a}} = 8 \left(\nabla_{[ \bar{d} |} G_{c11}{}^{c \bar{b}} + G_{c11}{}^{c \bar{b}} \nabla_{[\bar{d} |}\right) \phi_{\bar{b}| \bar{a}]} - 8 \left(\pd_{[\bar{d}} G_{\bar{a}]11}{}^{\bar{b}\bar{c}} + G_{[\bar{a}|11}{}^{\bar{b}\bar{c}} \nabla_{| \bar{d}]}\right) \phi_{\bar{b}\bar{c}} \, .
\ee 
The system (\ref{DelPhi}) consists of three coupled equations for three unknown functions. Any solutions $\phi_{\bar{a}\bar{b}}$ are determined by the flux only. In other words, the $\phi_{\bar{a}\bar{b}}$ zero modes do not depend on $\phi$. On the contrary, from the second equation in (\ref{211}) one can determine $\phi$ in terms of the flux and the solutions for $\phi_{\bar{a}\bar{b}}$. Hence, the number of solutions is determined by $\phi_{\bar{a}\bar{b}}$ and therefore for generic flux it is larger than in the fluxless case. The only way to reconcile this with the observations in the paragraph below the system (\ref{121}) is to assume that the solution is in fact $\phi_{\bar{a}\bar{b}} = 0$, which then implies that $\phi$ is harmonic. So the conclusion is that generic $(2,1,1)$ flux does not affect at all the zero modes.

Now let us turn to the $(1,2,1)$ type of flux. In this case the system (\ref{121}) reduces to:
\bea
\pd_{[\bar{a}} \phi_{\bar{b} \bar{c}]} + 4 G_{[\bar{a}\bar{b}}{}^{\bar{d}}{}_{|11|} 
\phi_{\bar{c}] \bar{d}} + 2 G^{\bar{d}}{}_{\bar{d}\,[\bar{a}\,|11|} \phi_{\bar{b}\bar{c}]} &=& 0 \nn \\
\pd_{\bar{a}} \phi + 4 g^{\bar{b} c} \pd_c \phi_{\bar{b} \bar{a}} + 
2 G_{\bar{a}\bar{c}}{}^{\bar{c}}{}_{11}\phi &=& 0 \, .
\eea
Acting with $\nabla^{\bar{a}}$ on the second equation, we find:
\be \label{phiev}
\Delta \phi - 4 G_{\bar{a}\bar{c}}{}^{\bar{c}}{}_{11} \pd^{\bar{a}}\phi - 4 (\pd^{\bar{a}} G_{\bar{a}\bar{c}}{}^{\bar{c}}{}_{11}) \phi = 0 \, .
\ee
The second-order linear differential operator acting on $\phi$ in the 
above equation is clearly elliptic. Since we are on a compact manifold, its 
spectrum will be discrete and so nontrivial solutions for $\phi$ will  
exist only if one of the eigenvalues is zero. This is clearly the case for vanishing flux, when the operator 
reduces to the laplacian and hence $\phi$ is harmonic. But for nonzero flux (unless the flux
is very particular) the eigenvalue will typically be shifted away from zero.
We conclude that, generically, the only solution of (\ref{phiev}) is $\phi=0$.
 In other words, generic flux completely lifts the zero modes of the fluxless case.

To recapitulate, turning on generic flux of type $(1,2,1)$ lifts all zero modes. 
Again, there may be important exceptions for very special choices of flux. On the other 
hand, a background flux component of type $(2,1,1)$ does not affect the zero mode counting, 
similarly to the $(2,2,0)$ component. We should note though that the situation is reversed 
for anti-M5-brane instantons. Namely, all zero modes on their worldvolume are lifted by a 
generic $(2,1,1)$ flux, whereas the $(1,2,1)$ type of flux does not affect them. For more 
details see the Appendix.

\section{World-volume flux} \label{WVfl}
\setcounter{equation}{0}

Until now, our considerations of the Dirac equation on the worldvolume of an M5-brane 
instanton always neglected for simplicity (as in all existing literature) the self-dual 
three-form $h$. However, as we saw in Section \ref{Res}, $h$ plays an important role in 
the resolution of the problem of reconciling five-brane instantons and gauged isometries. 
Hence, it is natural to ask how its presence affects the zero-mode counting of the 
previous sections. Unfortunately, taking into account both $h\neq 0$ (or equivalently, 
$H_{lmn} = (m^{-1})_l{}^p h_{mnp} \neq 0$) and nonvanishing background flux is too 
complicated to address in full generality. In the particular case of $(2,2,0)$ background 
though, the Dirac equation simplifies significantly and we will be able to analyze it in 
the presence of nonzero worldvolume flux. As a result, we will see that, whenever the 
topological constraints of Section \ref{Res} (or the conditions in Section \ref{NonSE}) are satisfied, the presence of M5-brane 
instantons is allowed by the zero-mode counting even with $h\neq 0$. 

\subsection{Preliminaries}

Since background fluxes of type $(2,2,0)$ (as those considered in Section \ref{reconc}) do not contribute to the Dirac equation (\ref{DEq}), for such backgrounds 
the latter simplifies to:
\be \label{Dirh}
\gamma_A m^{AB} \nabla_B \theta = 0 \, , \qquad m^{AB} = \delta^{AB} - 2 h^A{}_{CD} 
h^{BCD} \, .
\ee
To make further progress we will use the solution for $h$ found in \cite{MMMS} (see also 
\cite{BCN}):
\be \label{hsmall}
h = c \Omega + \chi \, .
\ee
Here $\Omega$ is the CY $(3,0)$-form, $\chi$ is a primitive $(1,2)$-form and $c$ is a 
constant, which for convenience we will absorb in the definition of $\Omega$ from now on. This is the most general form of the worldvolume flux for an M5 instanton 
wrapping a CY(3) in the absence of background flux.\footnote{Nonzero background flux 
complicates significantly the field equation for $h$ and the generic solution in that case 
is not known.} Nevertheless, it is all we need since effectively the compatibility condition between 
the background flux and M5 instantons is that the pullback of the flux on the brane 
worldvolume be zero (see Section \ref{Res}). Substituting (\ref{hsmall}) in equation 
(\ref{Dirh}) and using the decomposition (\ref{ThExp}) of the world-volume fermions, we 
find:
\bea \label{Dh}
\left( \pd_{\bar{a}} \phi + 4 g^{b \bar{c}} \pd_b \phi_{\bar{c} \bar{a}} - 
16 \mu^{\bar{c} \bar{b}} \nabla_{\bar{b}} \phi_{\bar{c}\bar{a}} - 4 \nu_{\bar{a}}{}^b 
\pd_b \phi \right) \gamma^{\bar{a}} | \Omega\rangle &=& 0 \nn \\
\left( \pd_{[\bar{a}} \phi_{\bar{b} \bar{c}]} - 4 \nu_{[\bar{a}}{}^d \pd_{|d|} 
\phi_{\bar{b}\bar{c}]} \right) \gamma^{\bar{a}\bar{b}\bar{c}}|\Omega\rangle &=& 0 \, ,
\eea
where for convenience we have introduced the combinations:\footnote{Our definition of 
$\mu_a{}^{\bar{b}}$ differs by a factor of $1/2$ from the one used in \cite{MMMS}.}
\be \label{nu2}
\nu_{\bar{a}}{}^b = \chi_{\bar{a}c\bar{d}} \chi^{bc\bar{d}} \, , \qquad \mu_a{}^{\bar{b}} 
= \Omega_{acd}\chi^{\bar{b}cd} \, ,
\ee
and used the relation $\mu^{\bar{a}\bar{b}} = \mu^{\bar{b}\bar{a}}$, whose origin will be recalled shortly.
Note that, inverting the second relation above i.e. using $\chi_{a\bar{b}
\bar{c}} = \mu_a{}^{\bar{d}} \bar{\Omega}_{\bar{d}\bar{b}\bar{c}}$, one can write 
$\nu_{\bar{a}}{}^b = \mu_c{}^{\bar{d}} \bar{\Omega}_{\bar{a}\bar{d}\bar{e}}
\mu_g{}^{\bar{e}} \bar{\Omega}^{gbc}$.

In order to be able to solve equations (\ref{Dh}), we will need one more result from 
\cite{MMMS}. Namely, the $(1,2)$-form $\chi$, that determines the world-volume flux via 
(\ref{hsmall}), has to be of a very particular form. Let us briefly recall the reasons for 
that. The self-dual three-form $h$ is determined by its equation of motion \cite{HSW}:
\be
m^{AB} \nabla_A h_{BCD} = 0 \, .
\ee
The $(1,1)$ and $(0,2)$ components of the latter give respectively
\be \label{KS}
\pd_{[a}\mu_{b]}{}^{\bar{c}} - 4c\mu_{[a}{}^{\bar{d}} \pd_{\bar{d}} \mu_{b]}{}^{\bar{c}} 
= 0 \, ,
\ee
which is exactly the Kodaira--Spencer equation \cite{KodSp} that describes finite 
deformations of the complex structure of the CY(3), and
\be \label{def}
\nabla^a \chi_{a\bar{b}\bar{c}} - 4 \chi_{\bar{a}d\bar{e}} \chi^{fd\bar{e}} 
\nabla^{\bar{a}} \chi_{f\bar{b}\bar{c}} = 0 \, ,
\ee
whereas the $(2,0)$ component vanishes identically due to $\nabla \Omega = 0$.
In addition, the primitivity condition $J\wedge \chi = 0$ leads to
\be \label{prim}
\mu_{ab} = \mu_{ba} \, .
\ee
Equation (\ref{def}) is a deformation of the gauge choice $\pd^{\dagger} \chi = 0$ in which 
the solution of (\ref{KS}) was found by Tian and Todorov \cite{TT}. This solution has the 
form
\be \label{chiser}
\chi = \sum_{n=1}^{\infty} \epsilon^n \chi^{(n)} \, ,
\ee
with $\epsilon$ being a small parameter, and satisfies (\ref{prim}) 
automatically.\footnote{This solution of the Kodaira--Spencer equation has also been 
considered in the context of the topological B--model in \cite{BCOV}.} As the 
precise form of the functions $\chi^{(n)}$ is not important for us, we will not 
write them down. It was argued in \cite{MMMS} that the Tian-Todorov solution can be 
deformed to a new one, still of the form (\ref{chiser}), which satisfies the gauge 
condition (\ref{def}) together with (\ref{KS}) and (\ref{prim}).\footnote{We should note, 
that although \cite{MMMS} presents convincing arguments, it does not give a rigorous proof.}
In view of this, we take the worldvolume flux parameters in the Dirac equation (\ref{Dh}) 
to also be power series in $\epsilon$:
\be
\mu = \sum_{n=1}^{\infty} \epsilon^n \mu^{(n)} \, , \qquad \nu = \sum_{n=1}^{\infty}
\epsilon^{2n} \nu^{(2n)} \, ,
\ee
where the expansion of $\nu$ has only even powers because of (\ref{nu2}). This implies 
that the solutions of (\ref{Dh}) should also be power series:
\be
\phi = \sum_{n=1}^{\infty} \epsilon^n \phi^{(n)} \, , \qquad \phi_{\bar{a}\bar{b}} = 
\sum_{n=1}^{\infty} \epsilon^n \phi_{\bar{a}\bar{b}}^{(n)} \, .
\ee

\subsection{Solving the Dirac equation}

Let us now start solving (\ref{Dh}) order by order in $\epsilon$. 
At first order we have the system:
\bea \label{Ep1}
\pd_{\bar{a}} \phi^{(1)} + 4 g^{b \bar{c}} \pd_b \phi^{(1)}_{\bar{c} \bar{a}} &=& 0 \nn \\
\pd_{[\bar{a}} \phi^{(1)}_{\bar{b} \bar{c}]} &=& 0 \, ,
\eea
which implies that $\phi^{(1)}$ and $\phi^{(1)}_{\bar{a}\bar{b}}$ are harmonic. Using 
$h^{0,2}(CY(3)) = 0$, we find that $\phi^{(1)}_{\bar{a}\bar{b}}=0$. Hence, at second order 
(\ref{Dh}) gives again:
\bea
\pd_{\bar{a}} \phi^{(2)} + 4 g^{b \bar{c}} \pd_b \phi^{(2)}_{\bar{c} \bar{a}} &=& 0 \nn \\
\pd_{[\bar{a}} \phi^{(2)}_{\bar{b} \bar{c}]} &=& 0 \, ,
\eea
since the only $\epsilon^2$ term containing $\mu$ or $\nu$ would have been 
$16 \mu^{(1)}{}^{\bar{c} \bar{b}} \nabla_{\bar{b}} \phi^{(1)}_{\bar{c}\bar{a}}$. 
Therefore, $\phi^{(2)}$ and $\phi^{(2)}_{\bar{a}\bar{b}}$ are also harmonic and as a 
result $\phi^{(2)}_{\bar{a}\bar{b}} = 0$ too. 

At order $\epsilon^3$ we find a more complicated system:
\bea \label{Ep3}
\pd_{\bar{a}} \phi^{(3)} + 4 g^{b \bar{c}} \pd_b \phi^{(3)}_{\bar{c} \bar{a}} - 4 
\nu^{(2)}{}_{\bar{a}}{}^b \pd_b \phi^{(1)} &=& 0 \nn \\
\pd_{[\bar{a}} \phi^{(3)}_{\bar{b} \bar{c}]} &=& 0 \, ,
\eea
where we have used the vanishing of $\phi^{(1)}_{\bar{a}\bar{b}}$ and 
$\phi^{(2)}_{\bar{a}\bar{b}}$. Note that these equations are of exactly the same form as 
(3.11) and (3.13) of \cite{KKT} with $4 \nu^{(2)}{}_{\bar{a}}{}^b \pd_b \phi^{(1)}$ 
playing the role of the inhomogeneous flux term. However, in our case things are even 
simpler as we have already found that $\phi^{(1)}$ is harmonic. Since a harmonic 
function on a compact space is necessarily constant, $\pd_b \phi^{(1)} = 0$. Therefore, 
we again find that $\phi^{(3)}$, $\phi^{(3)}_{\bar{a}\bar{b}}$ are harmonic and so 
$\phi^{(3)}_{\bar{a}\bar{b}} = 0$.

It is easy to generalize the above considerations to any order, but before doing that, let 
us gain more familiarity with the equations involved by writing down the systems that 
result for two more iterations. At order $\epsilon^4$ (\ref{Dh}) gives: 
\bea
\pd_{\bar{a}} \phi^{(4)} + 4 g^{b \bar{c}} \pd_b \phi^{(4)}_{\bar{c} \bar{a}} - 16 
\mu^{(1)}{}^{\bar{c} \bar{b}} \nabla_{\bar{b}} \phi^{(3)}_{\bar{c}\bar{a}} - 4 
\nu^{(2)}{}_{\bar{a}}{}^b \pd_b \phi^{(2)} &=& 0 \nn \\
\pd_{[\bar{a}} \phi^{(4)}_{\bar{b} \bar{c}]} &=& 0 \, ,
\eea
whereas at order $\epsilon^5$:
\bea 
\pd_{\bar{a}} \phi^{(5)} + 4 g^{b \bar{c}} \pd_b \phi^{(5)}_{\bar{c} \bar{a}} - 16 
\sum_{k=1}^2 \mu^{(k)}{}^{\bar{c} \bar{b}} \nabla_{\bar{b}} \phi^{(5-k)}_{\bar{c}\bar{a}} 
- 4 \sum_{k=1}^2 \nu^{(2k)}{}_{\bar{a}}{}^b \pd_b \phi^{(5-2k)} &=& 0 \nn \\
\pd_{[\bar{a}} \phi^{(5)}_{\bar{b} \bar{c}]} -4 \nu^{(2)}{}_{[\bar{a}}{}^d \pd_{|d|} 
\phi^{(3)}_{\bar{b}\bar{c}]} &=& 0 \, .
\eea
It is clear now that at order $\epsilon^n$ one has:
\bea \label{nthO}
\pd_{\bar{a}} \phi^{(n)} + 4 g^{b \bar{c}} \pd_b \phi^{(n)}_{\bar{c} \bar{a}} &=& 4 
T^{(n)}_{\bar{a}} \nn \\
\pd_{[\bar{a}} \phi^{(n)}_{\bar{b} \bar{c}]} &=& S^{(n)}_{\bar{a}\bar{b}\bar{c}} \, ,
\eea
where for convenience we have introduced the notation
\bea
T^{(n)}_{\bar{a}} &=& 4 \sum_{k=1}^{n-3} \mu^{(k)}{}^{\bar{c} \bar{b}} \nabla_{\bar{b}} 
\phi^{(n-k)}_{\bar{c}\bar{a}} + \sum_{k=1}^{[n/2]} \nu^{(2k)}{}_{\bar{a}}{}^b \pd_b 
\phi^{(n-2k)} \nn \\
S^{(n)}_{\bar{a}\bar{b}\bar{c}} &=& 4 \sum_{k=1}^{[(n-3)/2]} \nu^{(2k)}{}_{[\bar{a}}{}^d 
\pd_{|d|} \phi^{(n-2k)}_{\bar{b}\bar{c}]} \, ,
\eea
and the upper limits in the sums take into account that $\phi^{(1)}_{\bar{a}\bar{b}}, 
\phi^{(2)}_{\bar{a}\bar{b}} = 0$. Obviously $T^{(n)}_{\bar{a}}$ and 
$S^{(n)}_{\bar{a}\bar{b}\bar{c}}$ depend only on $\phi^{(k)}$ and 
$\phi^{(k)}_{\bar{a}\bar{b}}$ with $k<n$, which at the previous stages have been shown to 
be harmonic. The latter fact implies that $\phi^{(k)}_{\bar{a}\bar{b}} = 0$ and 
$\pd_b \phi^{(k)} = 0$, which leads to 
\be
T^{(n)}_{\bar{a}} = 0 \, , \qquad S^{(n)}_{\bar{a}\bar{b}\bar{c}} = 0 \, .
\ee 
Hence $\phi^{(n)}$ and $\phi^{(n)}_{\bar{a}\bar{b}}$ are also harmonic.

To recapitulate, the solution of (\ref{Dh}) is given by $\phi_{\bar{a}\bar{b}} = 0$ and 
$\phi$ -- harmonic. Since $h^{0,0} (CY(3)) = 1$, we find a single zero mode. Taking into 
account the ${\bf 4}$ index that we have suppressed for convenience, this means that 
there are four zero modes just as in the case without world-volume flux.

\section{Discussion}

In this work we considered the interplay between flux-induced gauging of isometries and M5-brane 
instantons in five-dimensional heterotic M-theory. We showed that the reconciliation of 
the above two competing effects is due to the enforcement of the Gauss' law on the 
instanton worldvolume. It occurs for CY 3-folds that satisfy certain topological 
constraints. We explained that these constraints are significantly eased by considering 
compactifications with nonstandard embedding. In addition, we investigated in detail the 
Dirac equation for the M5 worldvolume fermions in the presence of all possible types of 
supersymmetric background flux. It turned out that backgrounds of type $(2,2,0)$ and 
$(2,1,1)$ do not change the zero-mode counting of the fluxless case, whereas flux of type 
$(1,2,1)$ lifts all zero modes. (For anti-M5 instantons the roles of the $(2,1,1)$ and 
$(1,2,1)$ fluxes are reversed.) We also managed, for first time, to solve the Dirac 
equation with nonvanishing worldvolume flux, although under restricted conditions. 

In heterotic M-theory there is always background flux, as we recalled in the introduction. 
So it was indeed pressing to show the consistency of the gauged supergravity 
description when non-perturbative effects are taken into account. However, clearly one can 
turn on background flux in M-theory compactifications to five or four dimensions as well. 
While the four-dimensional case is important mostly in the context of moduli stabilization, 
the five-dimensional one is relevant also for the domain-wall/QFT correspondence \cite{Sken} and 
supersymmetric realizations of the Randall-Sundrum scenario \cite{RanSun}. With the latter motivation in 
mind, the work \cite{BG} studied M-theory compactifications to 5d with background flux and, 
in particular, derived the flux-induced gauging in the effective supergravity description\footnote{For more work on solutions in this theory see e.g. \cite{GSa}.}, 
similarly to the results of \cite{LM} for type II strings. It turns out that again the 
isometry of the universal hypermultiplet moduli space, that is given by constant shifts of 
the axionic scalar $\sigma$, is gauged by the flux.\footnote{For classification of all 
possible (irrespective of background flux) gaugings of the moduli space of the universal 
hypermultiplet see \cite{gauging}.} Hence the considerations of the present paper apply, pretty 
much literally, to this case as well. Finally, it is certainly of interest to also study, 
in the same vein as here, the zero mode counting on the worldvolume of membranes in 
M-theory flux compactifications to 4d as M2 instantons could contribute to the 
superpotential of the low-energy effective theory.

\section*{Acknowledgements}
We would like to thank A. Kashani-Poor for valuable correspondence and A. Lukas and D. 
Waldram for discussions. We are also grateful to the referee for insightful comments that helped improve the quality of the paper. The work of L.A. is supported by the EC Marie Curie Research 
Training Network MRTN-CT-2004-512194 {\it Superstrings}. K.Z. is supported by a PPARC 
grant ``Gauge theory, String Theory and Twistor Space Techniques''.

\appendix

\section{Zero-mode counting for anti-M5 instantons}
\setcounter{equation}{0}

In the main text we considered M5-branes and, in order to construct an explicit 
representation of the fermion states on their worldvolume, we defined the Clifford vacuum 
by $\gamma^a |\Omega\rangle = 0$. Then the states are obtained by acting on 
$|\Omega\rangle$ with the creation operators $\gamma^{\bar{a}}$. To represent the fermion 
states on an anti-M5-brane worldvolume one can define another Clifford vacuum 
$|\Omega'\rangle$ by 
\be
\gamma^{\bar{a}}|\Omega'\rangle=0\;.
\ee
Now the creation operators are $\gamma^a$ and so the decomposition of the worldvolume 
spinor $\theta'$ is: 
\be \label{Rep1}
\theta'=\phi'|\Omega'\rangle+\phi'_{ab}\gamma^{ab}|\Omega'\rangle\;.
\ee
This is in accord with the realization of anti-brane worldvolume states in string theory 
as complex conjugates of the corresponding brane states. Therefore, it is immediately 
obvious that anti-M5 instantons couple to the fluxes of type $(2,1,1)$ and $(1,2,1)$ in 
an opposite way compared to the M5 instantons. Hence it follows from the results of 
Section \ref{Otherfl} that the $(2,1,1)$ flux lifts all of their zero-modes, whereas the 
$(1,2,1)$ flux does not affect them.

It is worth noting that, unlike the case of anti-D-branes, for anti-M5-branes there is an alternative 
representation of the fermionic states. This is due to the fact that their worldvolume 
spinor $\theta'$ has definite chirality, which is correlated with the self-duality 
properties of the world-volume three-form $h$. More precisely, $\theta'$ is anti-chiral 
and so can be built as an expansion in terms of odd number of creation operators acting 
on the original vacuum $|\Omega\rangle$:
\be \label{Rep2}
\theta'=\phi_{\bar{a}}\gamma^{\bar{a}}|\Omega\rangle
+\phi_{\bar{a}\bar{b}\bar{c}}\gamma^{\bar{a}\bar{b}\bar{c}}|\Omega\rangle.
\ee
Clearly, if the two representations (\ref{Rep1}) and (\ref{Rep2}) are to describe the same 
physics, they have to be equivalent. And indeed they are, since one can write an explicit 
mapping between them:
\be
|\Omega'\rangle=\bar{\Omega}_{\bar{a}\bar{b}\bar{c}} \gamma^{\bar{a}\bar{b}\bar{c}}|\Omega\rangle \qquad {\rm and} \qquad \phi'=\Omega^{\bar{a}\bar{b}\bar{c}}\phi_{\bar{a}\bar{b}\bar{c}} \;\;\; , \;\;\; \phi'_{ab}={\Omega_{ab}}^{\bar{c}}\phi_{\bar{c}} \,\, ,
\ee
which is essentially the statement of Serre duality for the Calabi-Yau 3-fold.


\begin{thebibliography}{100}

\bibitem{DRS}
K. Dasgupta, G. Rajesh and S. Sethi, {\em M Theory, Orientifolds and G-Flux}, JHEP {\bf 9908} (1999) 023, hep-th/9908088.

\bibitem{GKP}
S.B. Giddings, S. Kachru and J. Polchinski, {\em Hierarchies from Fluxes in String 
Compactifications}, Phys. Rev. {\bf D66} (2002) 106006, hep-th/0105097.

\bibitem{Grana}
M. Gra\~{n}a, {\em Flux Compactifications in String Theory: A Comprehensive Review}, 
Phys. Rept. {\bf 423} (2006) 91, hep-th/0509003.

\bibitem{KKLT}
S. Kachru, R. Kallosh, A. Linde and S.P. Trivedi, {\em De Sitter Vacua in String Theory}, 
Phys. Rev. {\bf D68} (2003) 046005, hep-th/0301240.

\bibitem{KT}
A.-K. Kashani-Poor and A. Tomasiello, {\em A Stringy Test of Flux-Induced Isometry Gauging}, 
Nucl. Phys. {\bf B728} (2005) 135, hep-th/0505208.

\bibitem{HW}
P. Ho\v{r}ava and E. Witten, {\em Heterotic and Type I String Dynamics from Eleven Dimensions}, 
Nucl. Phys. {\bf B460} (1996) 506, hep-th/9510209; {\em Eleven-Dimensional Supergravity on 
a Manifold with Boundary}, Nucl. Phys. {\bf B475} (1996) 94, hep-th/9603142.

\bibitem{EW}
E. Witten, {\em Strong Coupling Expansion Of Calabi-Yau Compactification}, Nucl. Phys. 
{\bf B471} (1996) 135, hep-th/9602070.

\bibitem{BD}
T. Banks and M. Dine, {\em Couplings and Scales in Strongly Coupled Heterotic String Theory}, 
Nucl. Phys. {\bf B479} (1996) 173, hep-th/9605136.

\bibitem{cosm}
A. Lukas, B. Ovrut and D. Waldram, {\em Cosmological Solutions of Ho\v{r}ava-Witten Theory}, Phys. Rev. {\bf D60} (1999) 086001, hep-th/9806022; H. Reall, {\em Open and Closed Cosmological Solutions of Ho\v{r}ava-Witten Theory}, Phys. Rev. {\bf D59} (1999) 103506, hep-th/9809195; U. Ellwanger, {\em Cosmological Evolution in Compactified Ho\v{r}ava-Witten Theory Induced by Matter on the Branes}, Eur. Phys. J. {\bf C25} (2002) 157, hep-th/0001126; M. Braendle, A. Lukas and B. Ovrut, {\em Heterotic M-Theory Cosmology in Four and Five Dimensions}, Phys. Rev. {\bf D63} (2001) 026003, hep-th/0003256; C. van de Bruck, M. Dorca, R. Brandenberger and A. Lukas, {\em Cosmological Perturbations in Brane-World Theories: Formalism}, Phys. Rev. {\bf D62} (2000) 123515, hep-th/0005032; A. Kehagias and K. Tamvakis, {\em A Note on Brane Cosmology}, Phys. Lett. {\bf B515} (2001) 155, hep-ph/0104195; R. Arnowitt, J. Dent and B. Dutta, {\em Five Dimensional Cosmology in Ho\v{r}ava-Witten M-Theory}, Phys. Rev. {\bf D70} (2004) 126001, hep-th/0405050.

\bibitem{LOSW_DW}
A. Lukas, B. Ovrut, K. S. Stelle and D. Waldram, {\em The Universe as a Domain Wall}, Phys. Rev. {\bf D59} (1999) 086001, hep-th/9803235.

\bibitem{LOSW}
A. Lukas, B. Ovrut, K. S. Stelle and D. Waldram, {\em Heterotic M-theory in Five Dimensions}, 
Nucl. Phys. {\bf B552} (1999) 246, hep-th/9806051.

\bibitem{CKM5}
G. Curio and A. Krause, {\em $S$-$T$rack Stabilization of Heterotic de Sitter Vacua}, hep-th/0606243.

\bibitem{BB}
K. Becker and M. Becker, {\em Instanton Action for Type II Hypermultiplets}, Nucl. Phys. 
{\bf B551} (1999) 102, hep-th/9901126.

\bibitem{KKT}
R. Kallosh, A.-K. Kashani-Poor and A. Tomasiello, {\em Counting Fermionic Zero Modes on M5 
with Fluxes}, JHEP {\bf 0506} (2005) 069, hep-th/0503138.
 
\bibitem{NS}
N. Saulina, {\em Topological Constraints on Stabilized Flux Vacua}, Nucl. Phys. {\bf B720} 
(2005) 203, hep-th/0503125.

\bibitem{BKKST}
E. Bergshoeff, R. Kallosh, A.-K. Kashani-Poor, D. Sorokin and A. Tomasiello, {\em An Index 
for the Dirac Operator on D3 Branes with Background Fluxes}, JHEP 0510 (2005) 102, 
hep-th/0507069.

\bibitem{GKTT}
L. Goerlich, S. Kachru, P. K. Tripathy and S. P. Trivedi, {\em Gaugino Condensation and 
Nonperturbative Superpotentials in Flux Compactifications}, JHEP {\bf 0412} (2004) 074, 
hep-th/0407130.

\bibitem{BBSHM}
K. Becker, M. Becker and A. Strominger, {\em Fivebranes, Membranes and Non-Perturbative 
String Theory}, Nucl. Phys. {\bf B456} (1995) 130, hep-th/9507158; J. Harvey and G. Moore, 
{\em Superpotentials and Membrane Instantons}, hep-th/9907026.

\bibitem{WNonpSup}
E. Witten, {\em Non-Perturbative Superpotentials In String Theory}, Nucl. Phys. {\bf B474} 
(1996) 343, hep-th/9604030.

\bibitem{Vafa}
C. Vafa, {\em Evidence for F-Theory}, Nucl. Phys. {\bf B469} (1996) 403, hep-th/9602022.

\bibitem{KS}
R. Kallosh and D. Sorokin, {\em Dirac Action on M5 and M2 Branes with Bulk Fluxes}, JHEP 
{\bf 0505} (2005) 005, hep-th/0501081.

\bibitem{FS}
S. Ferrara and S. Sabharwal, {\em Dimensional Reduction of Type II Superstrings}, Class. 
Quantum Grav. {\bf 6} (1989) L77.

\bibitem{EWNewIss}
X. G. Wen and E. Witten, {\em World-sheet Instantons and the Peccei-Quinn Symmetry}, Phys. Lett. {\bf B166} (1986) 397.

\bibitem{AStr}
A. Strominger, {\em Loop Corrections to the Universal Hypermultiplet}, Phys. Lett. 
{\bf B421} (1998) 139, hep-th/9706195.

\bibitem{AMTV}
I. Antoniadis, R. Minasian, S. Theisen and P. Vanhove, {\em String Loop Corrections to 
the Universal Hypermultiplet}, Class. Quant. Grav. {\bf 20} (2003) 5079, hep-th/0307268.

\bibitem{ARV}
L. Anguelova, M. Ro\v{c}ek and S. Vandoren, {\em Quantum Corrections to the Universal 
Hypermultiplet and Superspace}, Phys. Rev. {\bf D70} (2004) 066001, hep-th/0402132.

\bibitem{dWvP}
B. de Wit and A. Van Proeyen, {\em Symmetries of Dual Quaternionic Manifolds}, Phys. Lett. 
{\bf B252} (1990) 221; B. de Wit, F. Vanderseypen and A. Van Proeyen, {\em Symmetry 
Structure of Special Geometries}, Nucl. Phys. {\bf 400} (1993) 463, hep-th/9210068.

\bibitem{SVandoren}
M. Davidse, U. Theis and S. Vandoren, {\em Fivebrane Instanton Corrections to the Universal Hypermultiplet}, Nucl. Phys. {\bf B697} (2004) 48, hep-th/0404147; M. Davidse, F. Saueressig, U. Theis and S. Vandoren, {\em Membrane Instantons and de Sitter Vacua}, JHEP {\bf 0509} (2005) 065, hep-th/0506097.

\bibitem{GS}
M. Gutperle and M. Spalinski, {\em Supergravity Instantons and the Universal Hypermultiplet}, JHEP {\bf 0006} (2000) 037, hep-th/0005068.

\bibitem{LLO}
Z. Lalak, A. Linde and B. Ovrut, {\em Soliton Solutions of M-theory on an Orbifold}, Phys. 
Lett. {\bf B425} (1998) 59, hep-th/9709214.

\bibitem{LOPR}
E. Lima, B. Ovrut, J. Park and R. Reinbacher, {\em Non-Perturbative Superpotentials from 
Membrane Instantons in Heterotic M-Theory}, Nucl. Phys. {\bf B614} (2001) 117, 
hep-th/0101049.

\bibitem{LRST}
D. L\"{u}st, S. Reffert, W. Schulgin and P. Tripathy, {\em Fermion Zero Modes in the Presence of Fluxes and a Non-perturbative Superpotential}, JHEP {\bf 0608} (2006) 071, hep-th/0509082.

\bibitem{BLNPST}
I. Bandos, K. Lechner, A. Nurmagambetov, P. Pasti, D. Sorokin, M. Tonin, {\em Covariant 
Action for the Super-Five-Brane of M-theory}, Phys. Rev. Lett. {\bf 78} (1997) 4332, 
hep-th/9701149; {\em On the Equivalence of Different Formulations of the M Theory 
Five-Brane}, Phys. Lett. {\bf B408} (1997) 135, hep-th/9703127.

\bibitem{HS}
P. Howe and E. Sezgin, {\em $D=11$, $p=5$}, Phys. Lett. {\bf B394} (1997) 62, 
hep-th/9611008.

\bibitem{HSW}
P. Howe, E. Sezgin and P. C. West, {\em The Six Dimensional Self-Dual Tensor}, Phys. Lett. 
{\bf B400} (1997) 255, hep-th/9702111.

\bibitem{nonst_weak}
E. Witten, {\em  New Issues In Manifolds Of Su(3) Holonomy}, Nucl. Phys. {\bf B268} (1986) 
79; L. Witten and E. Witten, {\em Large Radius Expansion Of Superstring Compactifications}, 
Nucl. Phys. {\bf B281} (1987) 109.

\bibitem{nonstemb}
K. Benakli, {\em Scales and Cosmological Applications of M Theory}, Phys. Lett. {\bf B447} 
(1999) 51, hep-th/9805181; Z. Lalak, S. Pokorski and S. Thomas, {\em Beyond the Standard 
Embedding in M-Theory on $S^1/Z_2$}, Nucl. Phys. {\bf B549} (1999) 63, hep-ph/9807503; R. 
Donagi, A. Lukas, B. Ovrut and D. Waldram, {\em Non-Perturbative Vacua and Particle 
Physics in M-Theory}, JHEP {\bf 9905} (1999) 018, hep-th/9811168; A. Lukas, B. Ovrut and 
D. Waldram, {\em Five--Branes and Supersymmetry Breaking in M--Theory}, JHEP {\bf 9904} 
(1999) 009, hep-th/9901017; Z. Lalak and S. Thomas, {\em Scales of Gaugino Condensation 
and Supersymmetry Breaking in Nonstandard M-Theory Embeddings}, Nucl. Phys. {\bf B575} 
(2000) 151, hep-th/9908147.

\bibitem{LOW}
A. Lukas, B. Ovrut and D. Waldram, {\em Non-standard Embedding and Five-branes in Heterotic 
M-theory}, Phys. Rev. {\bf D59} (1999) 106005, hep-th/9808101.

\bibitem{CK}
G. Curio and A. Krause, {\em Four-Flux and Warped Heterotic M-Theory Compactifications}, 
Nucl. Phys. {\bf B602} (2001) 172, hep-th/0012152; L. Anguelova and D. Vaman, {\em $R^4$ 
Corrections to Heterotic M-theory}, Nucl. Phys. {\bf B733} (2006) 132, hep-th/0506191.

\bibitem{AS}
A. Strominger, {\em Heterotic Solitons}, Nucl. Phys. {\bf B343} (1990) 167.

\bibitem{MMMS}
M. Marino, R. Minasian, G. Moore and A. Strominger, {\em Nonlinear Instantons from 
Supersymmetric p-branes}, JHEP {\bf 0001} (2000) 005, hep-th/9911206.

\bibitem{BCN}
L. Bao, M. Cederwall and B. E. W. Nilsson, {\em A Note on Topological M5-branes and 
String-Fivebrane Duality}, hep-th/0603120.

\bibitem{KodSp}
K. Kodaira and D. C. Spencer, {\em On Deformations of Complex Analytic Structures, I-II, 
III}, Annals of Math. {\bf 67} (1958) 328; Annals of Math {\bf 71} (1960) 43.

\bibitem{TT}
G. Tian, {\em Smoothness of the Universal Deformation Space of Compact Calabi-Yau 
Manifolds and its Petersson-Weil Metric}, in "Mathematical Aspects of String Theory", S. 
T. Yau, ed., World Scientific, 1987; A. Todorov, {\em The Weil-Petersson Geometry of the 
Moduli Space of $SU(n\ge 3)$ (Calabi-Yau) Manifolds. I}, Commun. Math. Phys. {\bf 126} 
(1989) 325.

\bibitem{BCOV}
M. Bershadsky, S. Cecotti, H. Ooguri and C. Vafa, {\em Kodaira--Spencer Theory of
Gravity and Exact Results for Quantum String Amplitudes}, 
Commun.Math.Phys. {\bf 165} (1994) 311, hep-th/9309140.

\bibitem{Sken}
H. J. Boonstra, K. Skenderis and P. K. Townsend, {\em The Domain-wall/QFT Correspondence}, JHEP {\bf 9901} (1999) 003, hep-th/9807137.

\bibitem{RanSun}
L. Randall and R. Sundrum, {\em  A Large Mass Hierarchy from a Small Extra Dimension}, Phys. Rev. Lett. {\bf 83} (1999) 3370, hep-ph/9905221; {\em  An Alternative to Compactification}, Phys. Rev. Lett {\bf 83} (1999) 4690, hep-th/9906064.

\bibitem{BG}
K. Behrndt and S. Gukov, {\em Domain Walls and Superpotentials from M Theory on Calabi-Yau 
Three-Folds}, Nucl.Phys. {\bf B580} (2000) 225, hep-th/0001082.

\bibitem{GSa}
M. Gutperle and W. Sabra, {\em A Supersymmetric Solution in N=2 Gauged Supergravity with the Universal Hypermultiplet}, Phys. Lett. {\bf B511} (2001) 311, hep-th/0104044.

\bibitem{LM}
J. Louis and A. Micu, {\em Type II Theories Compactified on Calabi-Yau Threefolds in the 
Presence of Background Fluxes}, Nucl. Phys. {\bf B635} (2002) 395, hep-th/0202168.

\bibitem{gauging}
K.~Behrndt and M.~Cveti\v{c}, {\em Gauging of $N=2$ Supergravity Hypermultiplet and Novel Renormalization Group Flows}, Nucl. Phys. {\bf B609} (2001) 183, hep-th/0101007; 
A.~Ceresole, G.~Dall'Agata, R.~Kallosh and A.~Van Proeyen, {\em Hypermultiplets, Domain Walls and Supersymmetric Attractors}, Phys. Rev. {\bf D64} (2001) 104006.

\end{thebibliography}
\end{document}